\documentclass[prb,aps,amsfonts,amssymb,floatfix,showpacs,twocolumn,superscripta
ddress]{revtex4}
\usepackage{graphicx}
\usepackage{dcolumn}% Align table columns on decimal point
\usepackage{bm}% bold math
\begin{document} %****
\title{Gutzwiller Magnetic Phase Diagram of the Undoped 
$t-t'-U$ Hubbard Model}
\author{R.S. Markiewicz}
\affiliation{ Physics Department, Northeastern University, Boston MA 
02115, USA}
\affiliation{SMC-INFM-CNR and Dipartimento di Fisica, Universit\`a di Roma
``La Sapienza'', P. Aldo Moro 2, 00185 Roma, Italy}
\affiliation{ISC-CNR, Via dei Taurini 19, I-00185 Roma, Italy}
\author{J. Lorenzana}
\affiliation{SMC-INFM-CNR and Dipartimento di Fisica, Universit\`a di Roma
``La Sapienza'', P. Aldo Moro 2, 00185 Roma, Italy}
\affiliation{ISC-CNR, Via dei Taurini 19, I-00185 Roma, Italy}
\author{G. Seibold}
\affiliation{Institut F\"ur Physik, BTU Cottbus, PBox 101344, 03013 Cottbus, 
Germany}
\begin{abstract}
We calculate the magnetic phase diagram of the half-filled $t-t'-U$ Hubbard 
model as a function of $t'$ and $U$, within the Gutzwiller approximation RPA (GA+RPA).  
As $U$ increases, the system first crosses over to one of a wide variety of incommensurate 
phases, whose origin is clarified in terms of double nesting. We evaluate the stability 
regime of the incommensurate phases by allowing for symmetry-breaking with regard to the 
formation of spin spirals, and find a crossover to commensurate phases as $U$ increases and 
a full gap opens. The results are compared with a variety of other recent calculations, and 
in general good agreement is found.  For parameters appropriate to the cuprates, double 
occupancy should be only mildly suppressed in the absence of magnetic order, inconsistent 
with a strong coupling scenario. 
\end{abstract} 
\maketitle

%\pacs{PACS numbers~:~~74.20.Mn,74.72.-h, 71.45.Lr, 74.50.+r }

%****
\narrowtext
%****

\section{Introduction}
An important issue in the Hubbard model is the nature of the metal-insulator transition, 
whether it is more Mott-like (driven by suppression of double occupancy, with no 
accompanying magnetic order) or more Slater-like (associated with magnetic order and a 
Stoner-factor instability).  The question is rather subtle, since for instance a Mott 
phase can have a significant exchange coupling, which can lead to a parasitic magnetic 
order at low temperatures.  Alternatively, critical fluctuations in a two-dimensional 
system will drive the magnetic ordering temperature to zero (Mermin-Wagner theorem) while 
leaving behind a finite temperature pseudogap.  Recently Tocchio, Becca, Parola, and 
Sorella (TBPS)\cite{TBPS,BTS} carried out a variational calculation of the $t-t'-U$ 
Hubbard model at half filling, and found that the whole $T=0$ phase diagram, Fig.~1(a) is 
dominated either by paramagnetic phases or by phases with long-range magnetic order 
(solid green line), except for a small window of nonmagnetic insulator (`spin liquid' -- 
dashed green line).  Here we show that the ordered 
magnetic phase boundaries can be well reproduced by simpler Gutzwiller calculations, and 
that all of these instabilities are only weakly renormalized from the RPA values.  These 
calculations are sufficiently simple that full allowance can be made for 
incommensurability [TBPS only studied antiferromagnetic order at $(\pi ,\pi )$ and 
$(\pi ,0)$], leading to a much richer phase diagram.  The domain where TBPS found the spin 
liquid phase is characterized by a large number of competing phases, leading to 
potential frustration.  

Using a Gutzwiller approximation (GA), Brinkman and Rice (BR)\cite{BR} found a sharp 
metal-insulator transition at a critical $U=U_{BR}$, where the effective mass diverges 
and the average double occupancy $n_d$ goes continuously to zero.  The BR line is 
$U_{BR}=8|E_k|$, where $E_k$ is the average kinetic energy per carrier below the Fermi 
energy $E_F$.  While the sharp second order transition is now known to be an artifact 
of the simplified variational scheme\cite{Faz}, $U_{BR}$ signals a crossover to a 
regime of small $n_d$, and hence can still serve as a measure of strong correlations.  

However, we will show in the present paper that the BR transition usually takes
place at much larger $U$ than a magnetic instability towards an incommensurate
magnetic state. We will give a detailed analyis of these instabilites in terms
of double nesting and discuss the differences between Hartree-Fock (HF) and the GA
approach concerning the magnetic phase diagram.
Before presenting the corresponding results in Sec. III we briefly introduce the
model and formalism in Sec. II and conclude our discussion in Sec. IV.

\begin{figure}
\includegraphics[width=8cm,clip=true]{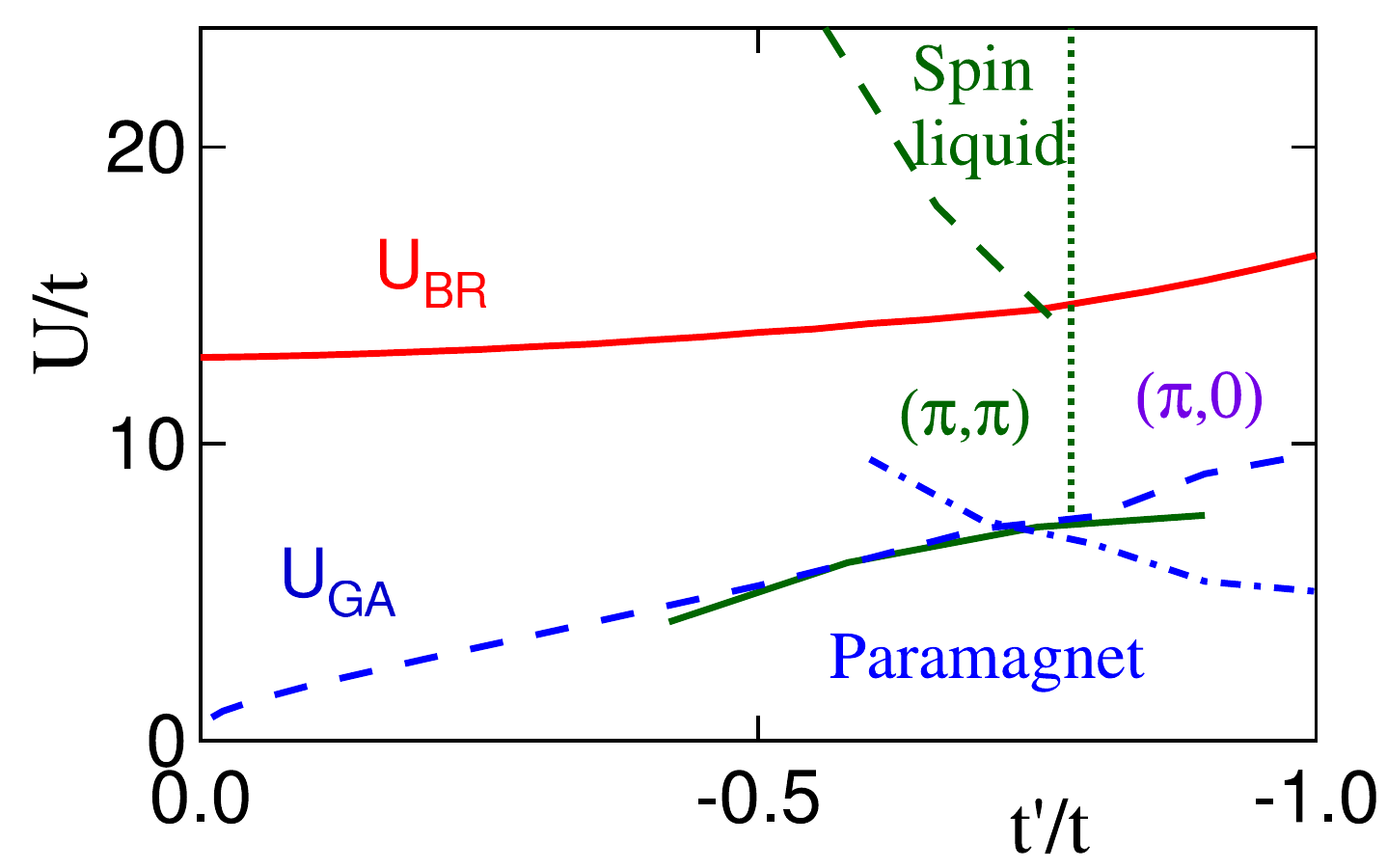}
\caption{(Color online.)
Phase diagram, as a function of $U$ and $t'$, showing $U_{BR}$ (red solid line), 
and calculations from Ref.~\protect\cite{TBPS} (green lines).  Shown also are GA+RPA
calculations  $\tilde U_{GA}$ in which the momentum of the instability is restricted to 
be $(\pi ,\pi )$ [dashed blue lines] or $(\pi ,0)$ [dot-dashed blue lines]. 
}
\label{fig:1}
\end{figure}

\section{Model and Formalism}
Starting point is the two-dimensional one-band Hubbard model
\begin{equation}\label{HM}
H= \sum_{i,j,\sigma} t_{ij} c_{i,\sigma}^{\dagger}c_{j,\sigma} + U\sum_{i}
n_{i,\uparrow}n_{i,\downarrow},
\end{equation}
where $c_{i,\sigma}$ ($c^\dagger_{i,\sigma}$) destroys (creates) an electron
with spin $\sigma$ at site
$i$, and $n_{i,\sigma}=c_{i,\sigma}^{\dagger}c_{i,\sigma}$. $U$ is the
on-site Hubbard repulsion and $t_{ij}$ denotes the hopping parameter between
sites $i$ and $j$. In the present paper we restrict to hopping between 
nearest ($\sim t$) and next-nearest ($\sim t'$)neighbors leading to a
dispersion in momentum space $\epsilon^0_{\bf k}=-2t[\cos(k_x)+\cos(k_y)]
-4t'\cos(k_x)\cos(k_y)$.

Our approach is based on a generalized GA \cite{GEBHARD}
supplemented with Gaussian fluctuations
(GA+RPA) \cite{goetz01} in order to evaluate the magnetic instabilities.
Since in the following we will also calculate spiral states 
we use a spin-rotational invariant
Gutzwiller energy functional  as derived e.g. in Ref. \onlinecite{jose06}

\begin{equation}\label{eq:ega}
E^{GA}= \sum_{i,j}
t_{ij} \langle{\bf \Psi_i}^\dagger {\bf z_{i}}
{\bf z_{j}}{\bf \Psi_j}\rangle + U\sum_{i} D_i  
\end{equation}
and $D_i$ denote the variational ('double occupancy') parameters.

We have also defined the spinor operators
\begin{displaymath}
{\bf \Psi_i}^\dagger = (c_{i\uparrow}^\dagger , c_{i\downarrow}^\dagger)
\,\,\,\,\,\,  {\bf \Psi_i} = \left(\begin{array}{c} c_{i\uparrow} \\
c_{i\downarrow} \end{array}\right)
\end{displaymath}
and  the ${\bf z}$-matrix

\begin{displaymath}
{\bf z}_i= \left( \begin{array}{cc}
z_{i\uparrow}\cos^2\frac{\varphi_i}{2}+z_{i\downarrow}
\sin^2\frac{\varphi_i}{2} &
\frac{S_i^-}{2S_i^z}[z_{i\uparrow}-z_{i\downarrow}]\cos\varphi_i \\
\frac{S_i^+}{2S_i^z}[z_{i\uparrow}-z_{i\downarrow}]\cos\varphi_i&
z_{i\uparrow}\sin^2\frac{\varphi_i}{2}+z_{i\downarrow}\cos^2\frac{\varphi_i}{2}
\end{array} \right)
\end{displaymath}
with
\begin{displaymath}
\tan^2\varphi_i=\frac{S_i^+S_i^-}{(S_i^z)^2}.
\end{displaymath}
In the limit of a vanishing rotation angle $\varphi$ the ${\bf z}$-matrix
becomes diagonal and the renormalization factors
\begin{widetext}
\begin{displaymath}
z_{i\sigma} = \frac{\sqrt{(1-\rho_i+D_i)(\frac{1}{2}\rho_i+\frac{S_i^z}
{\cos(\varphi_i)}-D_i)}+\sqrt{D_i(\frac{1}{2}\rho_i-\frac{S_i^z}
{\cos(\varphi_i)}-D_i)}}
{\sqrt{(\frac{1}{2}\rho_i+\frac{S_i^z}{\cos(\varphi_i)})(1-\frac{1}{2}\rho_i-
\frac{S_i^z}{\cos(\varphi_i)})}}
\end{displaymath}
\end{widetext}
reduce to those of the standard GA ($\equiv z_0$ for a paramagnetic system).
Spiral solutions are then computed by minimizing $E^{GA}$ with respect to
a homogeneous
rotation of spins with wave-vecor ${\bf Q}$
\begin{eqnarray}
S_i^x &=& S_0 \cos({\bf Q R}_i) \\
S_i^y &=& S_0 \sin({\bf Q R}_i).
\end{eqnarray}

In order to compute the magnetic instabilites 
one can derive an equation similar to the Stoner
criterion $U_{HF}=1/max\{\chi_0(q)\}$ in HF+RPA.
Here 
\begin{displaymath}
\chi_0(q)=-{1\over N}\sum_{{\bf k},\sigma}
{n_{{\bf k}+{\bf q},\sigma}-n_{{\bf k},\sigma}\over \epsilon^0_{{\bf k}+{\bf
q}}-\epsilon^0_{{\bf k}}}
\end{displaymath}
denotes the bare static susceptibility, and the maximum is taken over all 
$q$-values.

In order to derive the corresponding condition within the GA
one has to calculate the response of the system to an external
perturbation which couples to the spin degrees of freedom.
This can be achieved by expanding the energy functional
Eq. \ref{eq:ega} up to quadratic order in the (spin)density
fluctuations \cite{sei04} which yields the generalized GA Stoner
criterion
\begin{equation}
max\{U_{GA}\chi_0(q)\}=1,
\label{eq:1G}
\end{equation}
with an effective magnetic interaction
\begin{equation}
U_{GA}=\left\lbrack N_q+M_q[2\bar E_1+M_q(\bar E_1^2-\bar E_2^2)\chi_0/z_0^2]\right\rbrack
/z_0^2.
\label{eq:2G}
\end{equation}
The parameters 
$N_{\bf q}$ and $M_{\bf q}=z_0(z'-z'_{+-})$ are defined in the Appendix of 
Ref.\onlinecite{DLGS} and

\begin{equation}
\bar E_{i=1,2}=-{1\over N\chi_0}\sum_{{\bf k},\sigma}
(\epsilon^0_{{\bf k}+{\bf q}}+\epsilon^0_{{\bf k}})^i
{n_{{\bf k}+{\bf q},\sigma}-n_{{\bf k},\sigma}\over \epsilon^0_{{\bf k}+
{\bf q}}-\epsilon^0_{{\bf k}}}.
\label{eq:3G}
\end{equation}

As discussed in Refs. \onlinecite{MGu,DLGS} the
effective magnetic interaction $U_{GA}\le U$ and saturates 
as the bare $U\rightarrow\infty$
whereas in HF+RPA, $U$ can be arbitrarily large.
As a consequence, for a given momentum ${\bf q}$ HF+RPA yields a magnetic 
transition at any 
doping whereas the corresponding instabilities 
within the GA are usually confined to a specific doping range.

In the next section we compare our results with
those of TBPS which are based on a Gutzwiller variational calculation (without making 
the GA). In their approach magnetic phases are based on HF groundstates 
while the spin liquid phase is 
based on a BCS ground state.  The calculations go beyond simple Gutzwiller by including a 
Jastrow factor and a backflow correction.

%\leavevmode
%   \epsfxsize=0.45\textwidth\epsfbox{lscobbaBR6.eps}
%\vskip0.5cm

\begin{figure}
\includegraphics[width=8cm,clip=true]{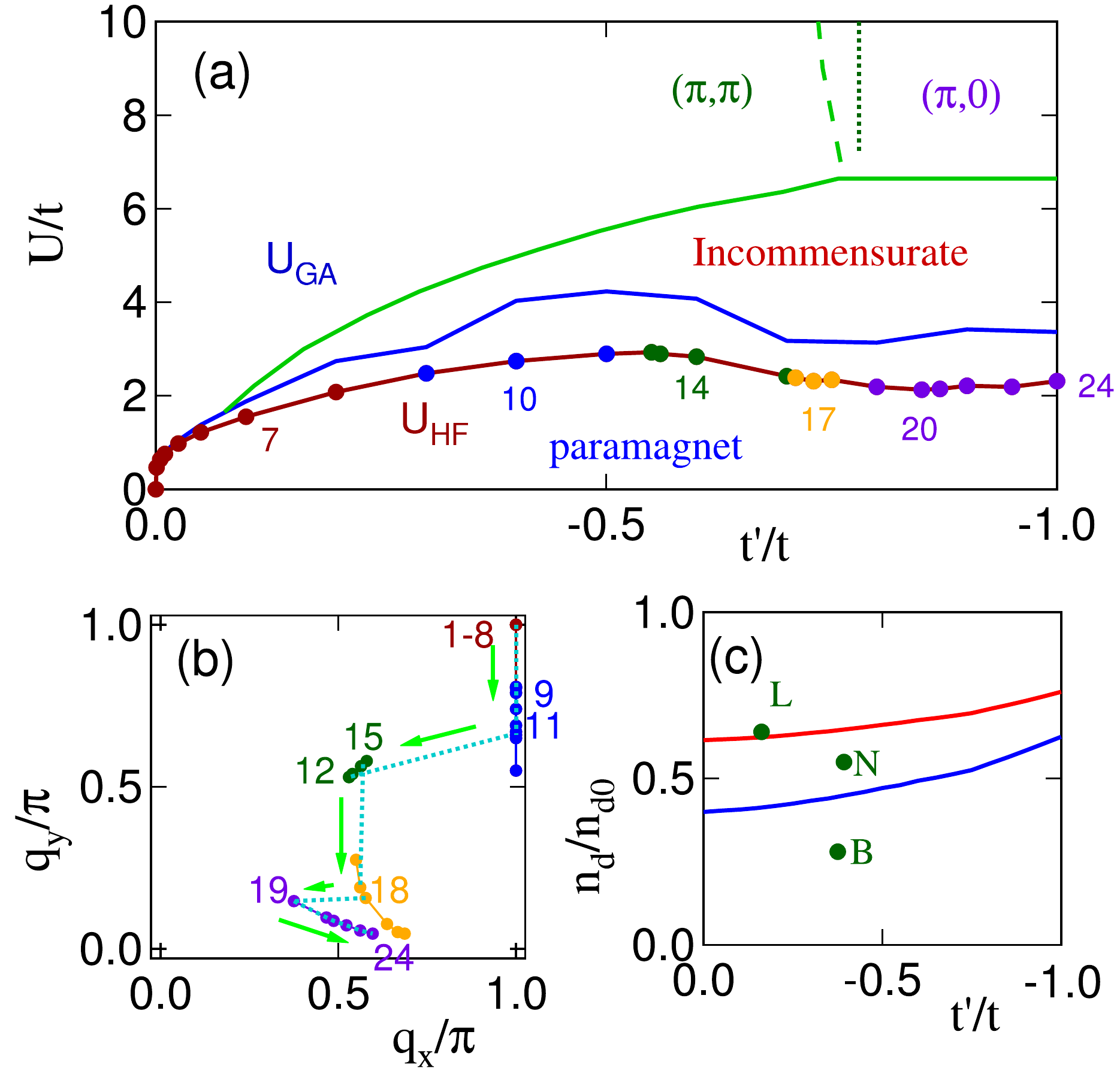}
\caption{(Color online.)
(a) Phase diagram obtained including the regime of incommensurate phases, comparing 
$U_{HF}$ (brown line) and $U_{GA}$ (blue line).  Shown also is the line of first order 
transitions to commensurate $(\pi ,\pi )$ or $(\pi ,0)$ order
(light green solid line) and the $(\pi ,\pi )$ to $(\pi ,0)$ crossover line (light 
green dashed line).  For ease in comparisons, the TBPS crossover line (green dotted 
line) from Fig.~1 is reproduced.  
To illustrate the dominant $q$-vector (Fig.~2(b)), the HF+RPA points are
color coded and numbered consecutively, although only some of the
numbers are indicated. 
(b) Position of the dominant susceptibility peak in HF+RPA, color code and numbers 
match the values in (a).  Note that some extra, metastable points are included, not 
shown in (a).  Blue dotted line traces evolution of stable points.
(c) $n_d/n_{d0}$ as a function of $t'$, assuming a constant $U=8t$ (red line) or
$10t$ (blue line).  Here $n_{d}$ is the double occupancy, and $n_{d0}=0.25$ is its 
uncorrelated value. Symbols indicate experimental dispersion renormalization
$Z_{disp}$ as a function of $t'$, from Ref.~\onlinecite{Arun3}, which should be an
experimental measure of $n_d/n_{d0}$.  Letters refer to L = La$_{2-x}$Sr$_x$CuO$_4$;
N = Nd$_{2-x}$Ce$_x$CuO$_{4\pm\delta}$; B = Bi$_2$Sr$_2$CaCu$_2$O$_8$.
}
\label{fig:1b}
\end{figure}
%\includegraphics[width=8cm,clip=true]{phasalpha.eps}
%\leavevmode
%   \epsfxsize=0.45\textwidth\epsfbox{lscobbaBR6b.eps}
%\vskip0.5cm

\section{Results}

Figure~1 compares the $t'/t - U$ phase diagram of TBPS\cite{TBPS,BTS} (green lines) with 
the Brinkman-Rice transition $U_{BR}$ (red solid line) and with the Gutzwiller transition 
$\tilde U_{GA}$, restricted to only $(\pi ,\pi )$ [$(\pi ,0)$] magnetic order 
(dashed [dotted] blue line).  It is seen that while the spin liquid phase falls above 
$U_{BR}$, magnetic phases arise for much lower $U$s, and the commensurate GA transitions 
are in excellent agreement with $U_{TBPS}$.  Since the phase boundaries depend only on 
$|t'|$, we illustrate only the situation $t'<0$ appropriate to the cuprates.

However, in general competing incommensurate phases become unstable first, due to Fermi 
surface nesting, Fig.~2(a), which compares the full $U_{GA}$ (blue line) with $U_{HF}$ 
(brown line).  While the HF+RPA calculation overestimates the stability of the magnetic 
phase, the overestimate is not very large.\cite{foot1}  Moreover, in all cases the most 
unstable $q$ vector (along these symmetry lines) is the same for the HF+RPA and the GA+RPA 
calculations.  Thus the main effect of the GA is to renormalize $U\rightarrow U_{GA}<U$, 
thereby reducing the range of the magnetic ordered phases. In Fig.~2(a) the HF+RPA 
calculations are coded with variously colored circles which match the points in Fig.~2(b), 
denoting the ordering $q$-vectors.  These changes are associated with the evolution of the 
Fermi surface (FS) with $t'$, as discussed below.  

Since the experimental $U$'s in cuprates fall in the range $\sim 6-8t$, Fig.~1 
suggests the cuprates are closer to Slater than to Mott physics. Within the BR model, 
the double occupancy at half filling is given by $n_d/n_{d0}=1- (U/U_{BR})^2=m/m^*$, 
where $n_{d0}=0.25$ is the uncorrelated double occupancy and $m^*$ the renormalized 
mass.  In Fig.~2(c) estimates for the double occupancy $n_d$ are plotted within the 
GA, for an assumed $U=8t$, $10t$, representative of the cuprates.  The relatively small 
reduction of $n_d$ explains why the HF+RPA results are so accurate.  Since $n_d\sim 
1/m^*$, the present results are consistent with the small observed enhancement of the 
effective mass (circles in Fig.~2(c))\cite{Arun3,foot2}.  Thus, in cuprates the 
double occupancy $n_d$ is reduced not by Mott physics but by strong AFM 
fluctuations.  In fact, the mean-field AFM suppresses double occupancy so well 
that Gutzwiller projection on an AFM ground state actually {\it increases} 
double occupancy.\cite{Faz}

In Fig.~2, it must be kept in mind that $U_{HF}$ and $U_{GA}$ define the onset of the 
magnetic instability, via a Stoner criterion.  As $U$ increases beyond the onset, a finite 
gap is opened and the optimal $q$ can change.  At large $U$ the entire Fermi surface is 
gapped, and the $q$'s which produce the largest gaps are favored.  

In order to access this crossover to commensurate phases we compute the energies of 
spiral textures within the spin rotational invariant extension of the GA 
\cite{fres92,sei04} as outlined in Sec. II. The most stable spiral is determined by
searching for the spiral wave-vector ${\bf Q}_{min}$ which minimizes the GA energy. 
For values slightly 
larger than $U_{GA}$ we consistently find the minimum of the energy landscape 
$E_{min}({\bf Q}_{min})$ close to the momenta of the instabilities as shown in Fig. 
\ref{fig:1b}a.  Upon further increasing $U/t$ and depending on $t'/t$ the momenta ${\bf 
Q}_{min}$ shift either towards the diagonal $(0,0) \to (\pi,\pi)$ or towards the line 
$(\pi,0)\to (\pi,\pi)$. For a critical $U/t$ one then finally finds a first order 
transition towards either  N\'{e}el order ${\bf Q}=(\pi,\pi)$ or towards linear 
antiferromagnetic (LAF) order at ${\bf Q}=(\pi,0)$ [or equivalently ${\bf Q}=(0,\pi 
)$].\cite{WhiSc}  This first order transition involves a {\it topological transition of 
the Fermi surface}, from a spiral phase with pockets to a fully gapped commensurate 
phase.

In Fig. \ref{fig:1b}a we show the first order line (light green) as 
an upper boundary for the incommensurate regime together with $U_{GA}$ as the lower 
transition (blue line) between incommensurate spin spirals and paramagnet.  The 
boundary between  N\'{e}el and LAF order (light green dashed line)  is close to the 
corresponding transition found by TBPS\cite{TBPS,BTS} (green dotted line).  Note that 
the (charge-rotationally invariant) GA samples all possible incommensuabilities, so we 
confirm that the two ordered phases found by TBPS are indeed the only allowed ordered 
phases for large enough $U$.  However, within the present scheme there is no energy 
gain for projected BCS wave-functions in the repulsive Hubbard model.  Therefore, while 
our simplified scheme allows for a detailed determination of magnetic phase
boundaries we cannot access the spin liquid regime found by TBPS (cf. \ref{fig:1b}c).
We also would like to  point out that besides spiral textures the incommensurate regime
may also contain  spin density wave solutions with an associated small charge density 
modulation.  However, for selected solutions we have checked that this has only a marginal 
influence on the first order transition line which in Fig. \ref{fig:1b} is of the order 
of the line width.

Figures~3 and~4 show how the FS evolves with $t'$, and how this is reflected in the peak 
susceptibility.  Fig.~3(a) displays a map of $\chi_0({\bf q})$ for $t'=-0.55t$, 
showing that $\chi_0$ is dominated by a complex series of ridges, partly defining a 
diamond-shaped plateau centered at $(\pi ,\pi )$.  The origin of these structures is 
readily apparent from Fig.~3(b) using the concept of nesting curves. 
For the generic case of two Fermi surface segments, a nesting curve can be defined as the 
locus of all points ${\bf q}={\bf k_{F1}}-{\bf k_{F2}}$, where ${\bf k_{Fi}}$ is a 
point on the $i$th FS, FS$_i$, with the restriction that when FS$_1$ is shifted by 
${\bf q}$ it is tangent to FS$_2$. For the parameters of Fig.~2(a,b)
one has one Fermi surface and the nesting curve is simply given by
${\bf q}=2{\bf k}_F$  where ${\bf k}_F$ is the 
(anisotropic) Fermi wave vector. The case of two Fermi surfaces is
discussed below. The dashed lines are extensions of the nesting curves 
folded back into the first BZ.  It can be seen that 
all of the sharp structure in $\chi_0$ falls along this nesting curve, and the susceptibility 
peaks correspond to points where two branches of the curve cross.  These values indicate 
$q$-vectors which nest the FS, and the crossing points correspond to double nesting, 
along two separate regions of the FS (see Fig.~4, below).

\begin{figure}
\includegraphics[width=8cm,clip=true]{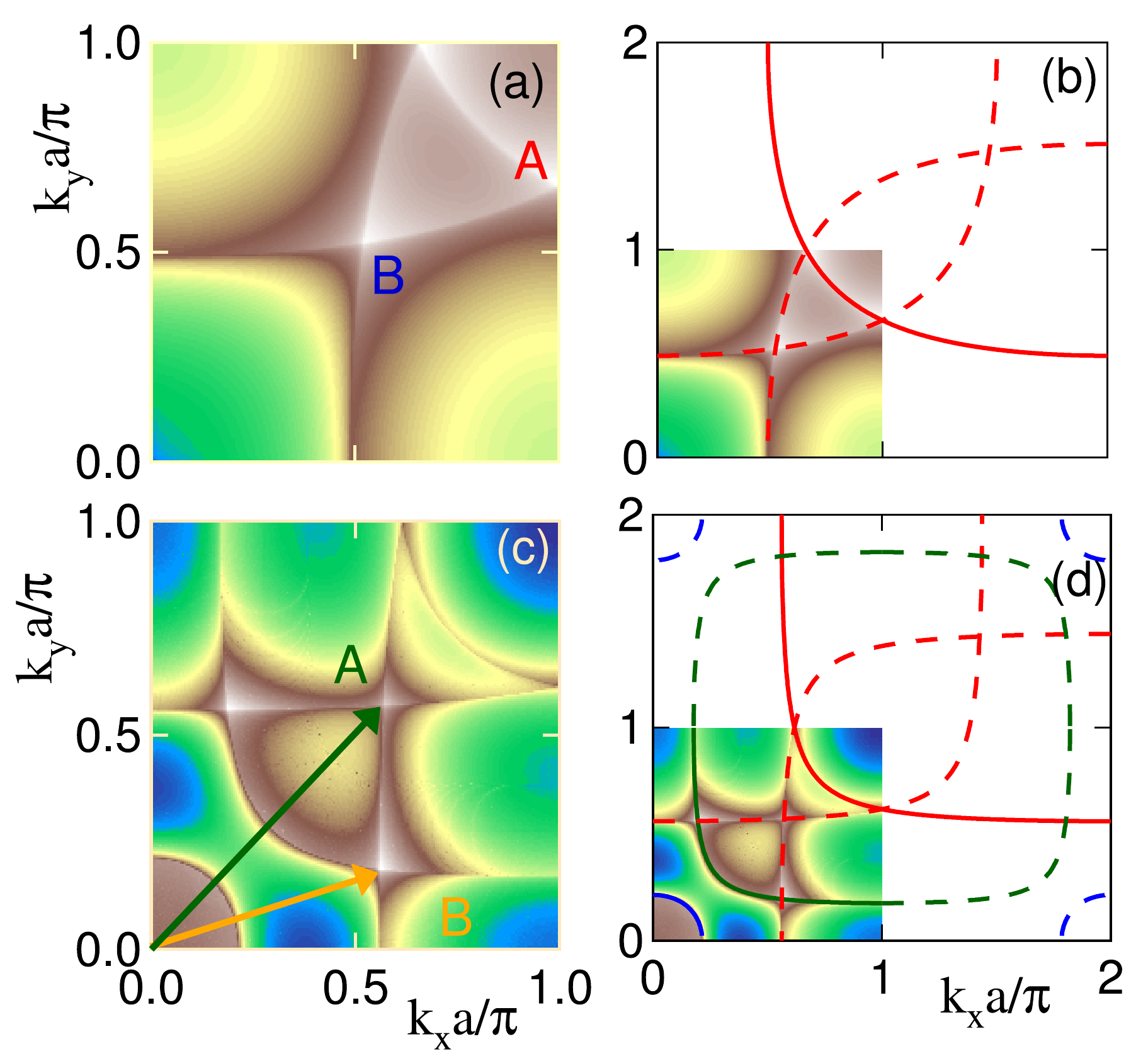}
%\leavevmode
%   \epsfxsize=0.45\textwidth\epsfbox{lscobbaBR3.eps}
%\vskip0.5cm
\caption{(Color online.)
(a) Plot of bare susceptibility $\chi_0$ in the first Brillouin zone, for 
$t'=-0.55$, at half filling.
(b) Similar plot with nesting curve (solid line) and its folded replicas (dashed 
lines).
(c,d) Similar plots for $t'=-0.73$, with additional susceptibility peaks and 
nesting curves, as discussed in text.
}
\label{fig:2}
\end{figure}

Figures~3(a,b) illustrate the typical behavior at small $|t'|$ (appropriate for cuprates).
From Fig.~2(a), the peak susceptibility is seen to be approximately commensurate at 
$(\pi ,\pi )$ for $|t'|\le 0.23t$ [brown circles], then becomes vertically 
incommensurate $(\pi -\delta ,\pi )$ for $0.23<|t'/t|\le 0.55$ [blue circles].  For 
$0.55<|t'/t|\le 0.72$, a diagonal incommensurate $(\pi -\delta ,\pi -\delta )$ phase 
arises [green circles].  Competition between these two phases, corresponding to the 
points $A$ and $B$ in Fig.~3(a), is also found in hole-doped cuprates\cite{MGu}.  
The positions of $A$ and $B$ are readily determined, Fig.~3(b).  Peak $A$ lies along 
the zone boundary ($q_xa=\pi$) with
\begin{equation}
q_ya =2arccos[{2t-E_F\over 2(t-2t')}],
\label{eq:1x}
\end{equation}
while peak $B$ follows the zone diagonal ($q_x=q_y$) with
\begin{equation}
q_xa =2arccos\Bigl[\sqrt{E_F\over 4t'}\Bigr].
\label{eq:1y}
\end{equation}
The only exception to this rule is the extended nearly-commensurate region near $(\pi 
,\pi )$ for small $|t'|$.

The topology of the band dispersion undergoes a drastic change at $|t'|=0.5t$.  At
that ratio the Van Hove singularity (VHS) moves down to the bottom of the band, leading
to a dispersionless band along the $x$ and $y$ axes.  Beyond $|t'/t|=0.5$, the VHS 
moves away from the band bottom, but now the point $\Gamma$ changes from the band 
minimum to a local maximum, opening the possibility of a second FS section.  At 
$|t'|=0.71t$, this section crosses the Fermi level at half filling, leading to a 
more complicated susceptibility map,
Fig.~3(c,d), with a second ${\bf q}=2{\bf k}_{F2}$ nesting curve (blue
line) centered on $\Gamma$, as  
well as a ${\bf q}={\bf k}_{F2}-{\bf k}_{F1}$ nesting curve (green
line).  In all cases the nesting curves  
match the positions of sharp structure in $\chi_0$.  [For the inter-FS nesting, the 
$q$-value corresponds to shifting one FS until it is tangent to the second.] This leads 
to a greatly increased number of nesting curve intersections, and the $\chi_0$ peak 
shifts, first, briefly, to a $k_{F1}$-mixed nesting curve intersection (orange line in 
Fig.~2(b)), then to a $k_{F2}$-mixed nesting curve (violet line).  Figure~3 provides an 
example of double nesting, showing two competing $q$-peaks, corresponding to $A$ and $B$ 
in Fig.~3(c).  Here nesting vector A [green arrows] involves double nesting of the large 
FS [green FSs], while in B [orange arrows] one nesting involves the large FS [brown FS], 
but the other involves nesting between the large FS and the $\Gamma$-centered 
pocket [orange FS].

\begin{figure}
\includegraphics[width=8cm,clip=true]{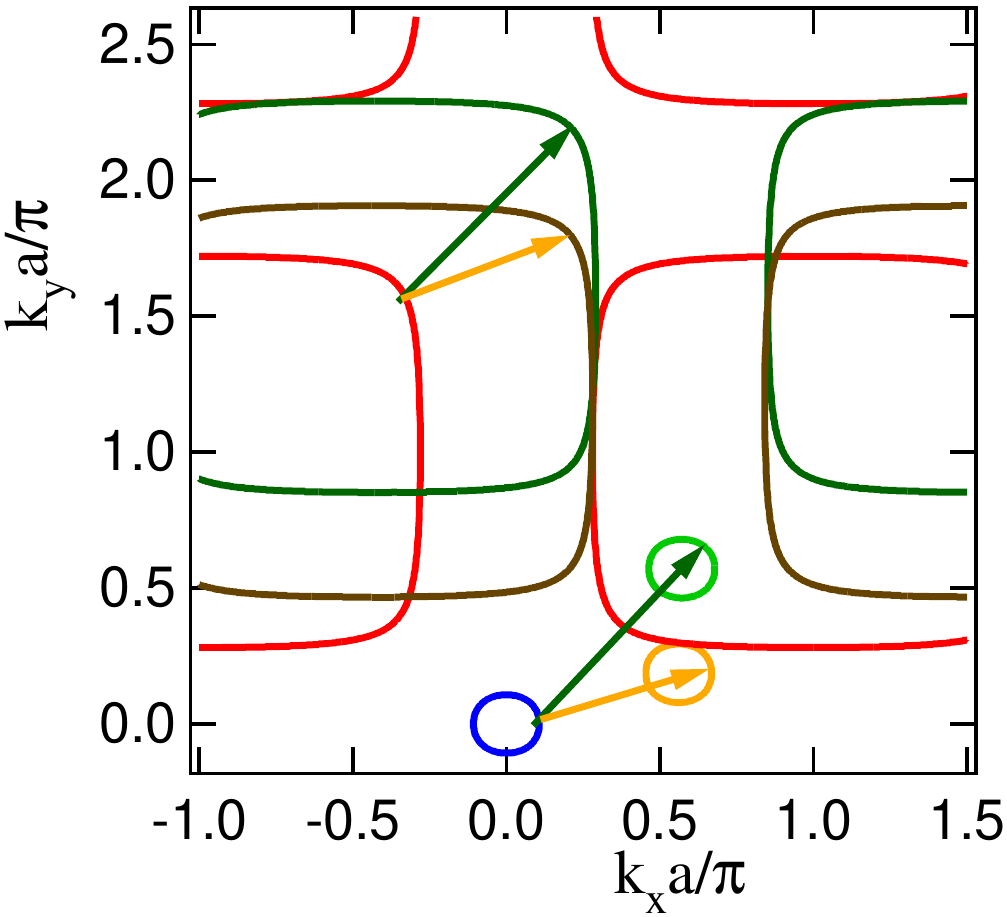}
%\leavevmode
%   \epsfxsize=0.40\textwidth\epsfbox{lscobbaBR4.eps}
%\vskip0.5cm
\caption{(Color online.)
Fermi surfaces (red and blue lines) and the $q$-shifted versions, 
illustrating two (competing) examples of double nesting.  Data are for 
$t'=-0.73t$, $x=0$, as in Fig.~2~(c,d) [arrows].}
\label{fig:3}
\end{figure}

%zzz
\section{Conclusions}
Increasingly, it is becoming clear that many of the complications of strongly correlated 
systems have to do with competing phases, whether leading to nanoscale phase separation, 
`stripes', or frustration.  However, part of the problem is a very incomplete 
understanding of phase competition even at weak coupling, and how the competing 
phases evolve between weak and strong coupling.  In this sense the magnetic 
instability of the two-dimensional Hubbard model provides an ideal case study, having a 
parameter space which is two-dimensional ($q_x$, $q_y$).  In HF+RPA $U$ is constant and 
the competition reduces to finding the maximum of $\chi_0$ (Stoner criterion).  We find 
that the Gutzwiller correction leads only to quantitative (close to numerical results) 
and not qualitative (gain/loss of new phase) changes.  Hence, the nesting 
curves introduced herein should provide valuable tools for determining the optimal 
nesting vectors for many different correlated materials, and how these evolve with doping, 
temperature, impurity scattering, etc.  

In commenting on the calculation of TBPS, Becca, Tocchio, and Sorella\cite{BTS} 
summarized earlier attempts to characterize the phase diagram of the model: 
``Remarkably, all these numerical approaches give very different results for the 
ground-state properties of this simple correlated model. In fact, there are huge 
discrepancies for determining the boundaries of various phases, but also for 
characterizing the most interesting non-magnetic insulator.''  Here, we have presented 
two additional phase diagrams, in HF and GA+RPA approximations.  A key point is that with
increasing U the first instability is generically to an incommensurate phase controlled by 
Fermi surface nesting.  For larger $U$ the FS is fully gapped, nesting is unimportant, and 
the only stable magnetic phases are the two commensurate phases found by TBPS.  For the 
rest our phase boundaries agree with TBPS except for the spin liquid phase which is beyond 
the capabilities of a simple mean field approximation.

%Finally, our calculations indicate that the cuprates are closer to Slater than to Mott physics.
%This is consistent with the observation of Ref. \cite{sahra} 
%that the bandstructure of LSCO is reasonably
%described within a standard band-theory picture up to binding energies of several hundred
%meV's. Correlation effects show up in the substantial loss of spectral weight of the metallic
%states with underdoping. Since the GA essentially is based on a renormalized conventional
%bandstructure this explains why in many cases (especially for broken-symmetry states) gives a 
%rather good account of ground state properties and and spectroscopic quantities \cite{lor02,lor03,
%sei05}.

%=========
%{\bf Acknowledgments}: 
\acknowledgements
This work is supported by the US Department of Energy, Office of  
Science, Basic Energy Sciences contract DE-FG02-07ER46352, and
benefited from the allocation of supercomputer time at NERSC,
Northeastern University's Advanced Scientific Computation Center
(ASCC).  RSM's work has been partially funded by the Marie Curie
Grant PIIF-GA-2008-220790 SOQCS, while GS' work is supported by
the Vigoni Program 2007-2008 of the Ateneo Italo-Tedesco
Deutsch-Italienisches Hochschulzentrum.


\begin{thebibliography}{99}
%\begin{references}
\bibitem{TBPS}L.F. Tocchio, F. Becca, A. Parola, and S. Sorella, 
Phys. Rev. B{\bf 78}, 041101(R) (2008). %cond-mat/0805.1476.
\bibitem{BTS}F. Becca, L.F. Tocchio, and S. Sorella, Proc. HFM2008 Conf., arXiv:0810.0665.
\bibitem{BR}W.F. Brinkman and T.M. Rice, Phys. Rev. B{\bf 2}, 4302 (1970).
\bibitem{Faz}P. Fazekas, ``Lecture Notes on Electron Correlation and Magnetism," 
(World Scientific, Singapore, 1999).
\bibitem{GEBHARD} F. Gebhard, Phys. Rev. B {\bf 41}, 9452 (1990).
\bibitem{goetz01} G. Seibold and J. Lorenzana, Phys. Rev. Lett. {\bf 86},
    2605 (2001).
\bibitem{jose06} J. Lorenzana and G. Seibold, Low Temp. Physics {\bf 32}, 320 (2006).
\bibitem{sei04}G. Seibold, F. Becca, P. Rubin, and J. Lorenzana, Phys. Rev. B {\bf 69}, 
155113 (2004).
\bibitem{WhiSc}C. Kusko and R.S. Markiewicz, Phys. Rev. B{\bf 65}, 041102(R) (2001).
\bibitem{MGu}R.S. Markiewicz, J. Lorenzana, G. Seibold, and A. Bansil, unpublished.
\bibitem{DLGS}A. Di Ciolo, J. Lorenzana, M. Grilli, and G. Seibold, Phys. Rev. B{\bf 
79}, 085101 (2009).  %arXiv:0806.3385.
\bibitem{KR}G. Kotliar and A.E. Ruckenstein, Phys. Rev. Lett. {\bf 57}, 1362 (1986).
\bibitem{foot1}Part of this difference is due to the fact that the RPA calculation samples 
all $q$ values (Fig.~2(a), while the Gutzwiller calculation was restricted to the high 
symmetry axes, $\Gamma\rightarrow (\pi ,0)\rightarrow (\pi ,\pi )\rightarrow\Gamma$.  
\bibitem{Arun3}R.S. Markiewicz, S. Sahrakorpi, M. Lindroos, Hsin Lin, and A. Bansil,
Phys. Rev. B{\bf 72}, 054519 (2005).  In Fig.~2(c) we introduce an effective 
$t'_{eff}=t'-t"$, which controls the energy of the Van Hove singularity.
\bibitem{foot2}While $m^*$ is determined at finite doping, for $U/U_{BR}\sim 0.6$ [$U\sim 8t$]
$m^*$ is not expected to vary greatly with doping\cite{Faz}.
\bibitem{fres92}R. Fr\'{e}sard and P. W\"olfle, J. Phys. Condens. Matter {\bf 4}, 3625 
(1992).
\bibitem{sahra} S. Sahrakorpi, R. S. Markiewicz, Hsin Lin, M. Lindroos, X. J. Zhou, T. Yoshida,
W. L. Yang, T. Kakeshita, H. Eisaki, S. Uchida, Seiki Komiya, Yoichi Ando, F. Zhou, Z. X. Zhao,
T. Sasagawa, A. Fujimori, Z. Hussain, Z.-X. Shen, and A. Bansil, Phys. Rev. B{\bf 78}, 104513 (2008).
%\bibitem{lor02} J. Lorenzana and G. Seibold, Phys. Rev. Lett. {\bf 89},
%                136401 (2002).
%\bibitem{lor03} J. Lorenzana and G. Seibold, Phys. Rev. Lett. {\bf 90},
%                66404 (2003).
%\bibitem{sei05} G. Seibold and J. Lorenzana, Phys. Rev. Lett. {\bf 94},
%                107006 (2005).

%\end{references}
\end{thebibliography}
\end{document}